\documentclass[aps,prd,preprint,nofootinbib]{revtex4-1}
\usepackage{amsmath,amssymb,graphics,graphicx,color}
\usepackage{scrextend}
\usepackage{hyperref}
\usepackage{float}
\usepackage{xcolor}
\usepackage[utf8]{inputenc}
\usepackage[english]{babel}

\begin{document}

\title{Sensitivities on dark photon from the Forward Physics Experiments}
\author{Kingman Cheung$^{a,b,c}$, C.J. Ouseph$^{a,b}$}
\affiliation{
  $^a$ Department of Physics, National Tsing Hua University, Hsinchu 30013,
  Taiwan\\
  $^b$ Center for Theory and Computation, National Tsing Hua University, Hsinchu 30013, Taiwan \\
  $^c$ Division of Quantum Phases and Devices, School of Physics,
  Konkuk University, Seoul 143-701, Republic of Korea
}
\date{\today}
\begin{abstract}
Neutrino-electron scattering experiments can explore the potential presence of a light gauge boson $A'$ which arises from an additional $U(1)_{B-L}$ group, or a dark photon $A'$ which arises from a dark sector and has kinetic mixing with the SM hypercharge gauge field.
We generically call it a dark photon.
In this study, we investigate the effect of the dark photon on neutrino-electron scattering $\nu e^-\rightarrow\nu e^-$ at the newly proposed forward physics experiments such as FASER$\nu$, FASER$\nu2$, SND@LHC and FLArE(10 tons).  We estimate the anticipated sensitivities to the $U(1)_{B-L}$ gauge coupling in a wide range of the dark photon mass $M_{A'}$. We compare the sensitivities of the proposed forward physics experiments with the current limits from TEXONO, GEMMA, BOREXINO, LSND, and CHARM II as well as NA64e experiments. We also 
extend the calculation to obtain the sensitivities on the kinetic mixing parameter $\epsilon$ in a wide range of dark photon mass $M_{A'}$. We demonstrate that the sensitivities do not improve for $M_{A'} < 1 $ GeV 
at the Forward Physics Facilities.
\end{abstract}
\maketitle

\section{Introduction}

Neutrino-charged-lepton scattering and neutrino-nucleon scattering have a 
long history of establishing the standard model (SM) \cite{Workman:2022ynf},
e.g., the discovery of neutral current in $\nu N$ scattering in 1973 
\cite{Freedman:1973yd}, the measurement of the weak-mixing angle, as well as constraining new physics that violates parity \cite{Workman:2022ynf}.
In the last two decades, neutrino physics mainly concerned with 
the oscillations of flavors of neutrinos and now the oscillation parameters
are reasonably well determined, except for the CP-violation parameter, 
the mass hierarchy of neutrino masses and $\theta_{23}$-octant degeneracy.
A number of new experiments are proposed to solve these subtleties, such as
DUNE, JUNO, Hyper-K, etc. 

On the other hand, neutrino scattering received renew interests 
in a number of proposals, including the near detector of DUNE making use of an off-axis neutrino beam, and FASER$\nu$ making use of the neutrinos
coming from the interaction point (IP) of the ATLAS experiment. FASER$\nu$ is indeed running and taking data \cite{talk-2022,talk-2022(2)} 
The most distinct feature of the neutrinos coming off the IP of ATLAS is that the 
energy range can be as high as TeV, which provides an unprecedented 
energy scale of studying neutrino-nucleon and neutrino-electron scattering.
The FASER$\nu$ experiment can cover a unique energy range that other 
experiments cannot cover. The ICECUBE focuses on very high-energy neutrinos with energy 10 TeV to 1 PeV. On the other hand, the short- and long-baseline experiments cover mostly around MeV up to a few GeV. There are no precise measurements of neutrino scattering in a few tens of GeV to a few hundreds of GeV region. FASER$\nu$ based on the neutrino flux coming off the LHC opens such a unique window in this energy range.
A further proposal is to build the Forward Physics Facilities (FPF)~\cite{Feng:2022inv} that can house a number of proposed experiments 
within the same location, including FASER2, Advanced SND, and FORMOSA. 
Thus, such experiments open an avenue to search for new physics associated 
with the neutrinos, which are not easily tested at the $pp$ collisions of 
the LHC experiments.

In this work, we investigate the potential of detecting a light $U(1)_{B-L}$
gauge boson or a hidden $U(1)$ gauge boson (aka. dark photon) at such proposed neutrino-scattering experiments. 
Generically, a hidden $U(1)$ does not couple to the SM particles, but it
can mix with the SM $U(1)_Y$ gauge field through the $B^{\mu\nu} F'_{\mu\nu} $ 
term.  Throughout the work we use the $U(1)_{B-L}$ gauge boson as illustration
and at the end we also extend the calculation to obtain the sensitivities on the kinetic
mixing $\epsilon$.

The organization is as follows. In the next section, we describe the 
$U(1)_{B-L}$ model and the relation to the hidden $U(1)$ model 
(dark photon). In Sec. III, we calculate the neutrino-electron scattering
in the SM and the $U(1)_{B-L}$ model. In Sec. IV, we estimate the sensitivity
reach at FASER$\nu$, FASER$\nu 2$, SND@LHC, and FLArE. We compare the 
results of Sec. IV to other experiments in Sec. V. We conclude in Sec. VI.
 
\section{The $U(1)_{B-L}$ Model}

We first introduce the $U(1)_{B-L}$ gauge interactions of a new gauge boson, denoted by $A'$, to the SM fermions with the charge proportional to quantum
number $B-L$ of the fermions, where $B$ and $L$ are the baryon and lepton
number, respectively. The interaction Lagrangian is given by 
\begin{equation}
\label{Eq.1}
     \mathcal{L}_{B-L}\supset g_{B-L} 
      \biggr [ - \bar{l}\gamma^{\mu}A'_{\mu}l
               -\bar{\nu}_\alpha\gamma^{\mu}A'_{\mu}\nu_{\alpha}
               + \frac{1}{3}\bar{q}\gamma^{\mu}A'_{\mu}q \biggr ]\;,
 \end{equation}
where $g_{B-L}$ is the $U(1)_{B-L}$ coupling strength and $q, l$ and $\nu$ are quark, charged lepton and neutrino fields, respectively. 
Here the gauge boson $A'$ couples vectorially to the SM fermions and the gauge boson $A'$ receives the mass by either a 
spontaneous breaking of the $U(1)_{B-L}$ or Stueckelberg mechanism, which
is not our concern here. Thus, the model is characterized by the mass 
$M_{A'}$ and the gauge coupling strength $g_{B-L}$.

Hidden-sector $U(1)'$ models belong to a class of hidden-valley models
\cite{Strassler:2006im}. It is often used to build models of dark matter.
Such $U(1)'$ models can be connected to the SM via a small kinetic mixing 
though it is hidden from our world. The mass of $A'$ and mixing parameter
$\epsilon$ can be tuned to satisfy the relic density of the Universe.
The Lagrangian describing the kinetic mixing with the SM hypercharge field
is given by
\begin{equation}\label{Eq.2}
    \mathcal{L}=-\frac{1}{4}B'^2_{\mu\nu}
      -\frac{1}{4}F''^2_{\mu \nu}+\frac{1}{2}\epsilon 
       B'_{\mu\nu}  F''^{\mu\nu}+\frac{1}{2} M^2_{A'}A''^2_{\mu} 
       - g_Y \frac{Y}{2} B'_\mu \bar f \gamma^\mu f \,,
\end{equation}
where $B'_{\mu\nu} = \partial_\mu B'_\nu - \partial_\nu B'_\mu $ and 
$F''_{\mu\nu} = \partial_\mu A''_\nu - \partial_\nu A''_\mu $, and here
the primed fields $B'$ and $A''$ denote the fields before rotating into physical fields.
The last term denotes the interaction of $B'_\mu$ with SM fermions.
The kinetic  mixing term can be eliminated by rotating $(B'_\mu, A''_\mu)$
into physical fields $(B_\mu, A'_\mu)$ by a field redefinition:
\begin{eqnarray}
   B'_\mu &\simeq & B_\mu + \epsilon A'_\mu  \nonumber \\
   A''_\mu & \simeq & A'_\mu \nonumber    \,.
\end{eqnarray}
Then the Lagrangian becomes
\begin{equation}\label{Eq.3}
    \mathcal{L}=-\frac{1}{4}B^2_{\mu\nu}
      -\frac{1}{4}F'^2_{\mu \nu}+ 
        \frac{1}{2} M^2_{A'}A'^2_{\mu} 
       - g_Y \frac{Y}{2} ( B_\mu + \epsilon A'_\mu)\, \bar f \gamma^\mu f \,.
\end{equation}
Thus, the hidden $U(1)'$ gauge boson $A'$ can now couple to the SM fermion
with an interaction strength $ \epsilon g_Y (Y/2)$. 

Conventionally, the term "dark photon" is used to refer to a $U(1)'$ gauge boson that is coupled to the SM only through kinetic mixing with the photon. 
Strictly speaking such a dark photon does not couple to neutrinos because of 
zero electric charge of the neutrino. However, as shown above the kinetic mixing
with the $B_\mu$ does create an interaction of $A'$ 
with neutrinos proportional to
the hypercharge of the neutrinos. Simultaneously, it also mixes with the SM
$Z$ boson and thus the mixing parameter $\epsilon$ is constrained to 
be small by the LEP data \cite{Workman:2022ynf}. Nevertheless,
we will continue to use the term dark photon in a more general sense for 
the gauge bosons coupled to SM particles only through the kinetic mixing.

The $U(1)_{B-L}$ gauge boson couples to the left- and right-handed fermions 
with the same quantum number, and thus the interaction is vector-like. 

On the other hand, the hidden $U(1)'$ gauge boson discussed above
couples to SM fermions with quantum number equal to the hypercharge of 
the fermion: $- \epsilon g_Y (Y/2) \bar f \gamma^\mu f \, A'_{\mu}$. 
Thus, for the left-handed electron and neutrino the couplings are
$- \epsilon g_Y (-1/2) \bar \nu \gamma \nu A'_\mu$ and
$- \epsilon g_Y (-1/2) \overline{e_L} \gamma e_L  A'_\mu$, respectively, 
while the coupling of the right-handed electron is
$- \epsilon g_Y (-1) \overline{e_R} \gamma e_R  A'_\mu$.

\section{Neutrino-Electron scattering}

Neutrino-electron scattering is a viable alternative to collider searches for physics beyond the SM
\cite{Feng:2022inv}.
In this section, we discuss the neutrino-electron scattering 
in the SM and with the presence of a dark photon.

\subsection{Standard Model cross section}
\begin{table}[ht]
\caption{\label{tb.1}\small Valued for $a$ and $b$ in the expression of 
Eq.~(\ref{Eq.5}), where $\alpha=\mu,~\tau$}
	\begin{tabular}{c c c c c}
		\hline\hline
	Process ~~~~&$a$&~~~~$b$\\
		[0.5ex] 
		\hline
		$\nu_{e} e^-\rightarrow \nu_e e^-$&~~~$\sin^2\theta_w+\frac{1}{2}$&~~~~$\sin^2\theta_w$\\
		$\bar{\nu}_{e} e^-\rightarrow \bar{\nu}_e e^-$&~~~$\sin^2\theta_w$&~~~~$\sin^2\theta_w+\frac{1}{2}$\\
		$\nu_{\alpha} e^-\rightarrow \nu_\alpha e^-$&~~~$\sin^2\theta_w-\frac{1}{2}$&~~~~$\sin^2\theta_w$\\
		$\bar{\nu}_{\alpha} e^-\rightarrow \bar{\nu}_\alpha e^-$&~~~$\sin^2\theta_w$&~~~~$\sin^2\theta_w-\frac{1}{2}$\\
		\hline
		\end{tabular}
\end{table}

In SM, the $\nu_e$-$e$ scattering takes place via the charged and neutral current interactions. However, the $\nu_\mu$-$e$ and $\nu_\tau$-$e$ scatterings are mediated only by the neutral $Z$ boson. The differential cross-section in the rest frame of the electron can be expressed as \cite{Bilmis:2015lja,Chakraborty:2021apc}
\begin{equation}\label{Eq.5}
\big[\frac{d\sigma}{dT}(\nu e^-\rightarrow \nu e^-)\big]_{SM}=\frac{2G^2_{\rm F}m_{\rm e}}{\pi E^2_{\rm \nu}}(a^2E^2_{\rm \nu}+b^2(E_{\rm \nu}-T)^2-abm_{\rm e}T) \;,
\end{equation}
where $G_{\rm F}$ denotes the Fermi constant, $T$ indicates the recoil energy of the electron, $E_{\rm \nu}$ is the energy of the incoming neutrino, and $m_{\rm e}$ is the mass of electron. The values for $a$ and $b$ depend on the neutrino flavor, which are given in Table. \ref{tb.1}. 
The contributing Feynman diagrams of the SM neutrino-electron scattering in the SM are shown in Fig. \ref{fig0}(a).

The SM production of $\nu_{e,\mu,\tau} e^- \to \nu_{e,\mu,\tau}
    e^- $ is the irreducible background to the scattering via the dark
    photon. On the other hand, a reducible background is the
    charged-current scattering with the nucleon $\nu_e N \to e^- N'$.
    In the signal process, the incoming neutrino does not scatter
    with the nucleon, and therefore the nucleon remains intact. On the
    other hand, the charged-current scattering with the nucleon will
    break the nucleon and thus distinguishable from the signal
    process.
     Nevertheless, the hadronic part in this charged current interaction of $\nu$ with the nucleon may not be fully identified with a 100\% efficiency. Therefore, whether this reducible background can be completely removed depends on the experimental setup and resolution.  This is beyond the scope of this work. To a large extent we can include this uncertainty into the systematic uncertainties. We will show the results with systematic uncertainties of 5\% and 20\%, the latter should be large enough to account for the uncertainty of this reducible background.

\subsection{Dark photon contribution to the cross-section }

The $U(1)_{B-L}$ gauge boson has vectorial couplings to SM fermions,
but the $U(1)'$ gauge boson has different couplings to left- and right-handed
fermions. In the following, we focus on the $U(1)_{B-L}$.
Later, we extend the calculation to the $U(1)'$.

\begin{figure}[ht!]
	\centering
	\includegraphics[width=7cm,height=4cm]{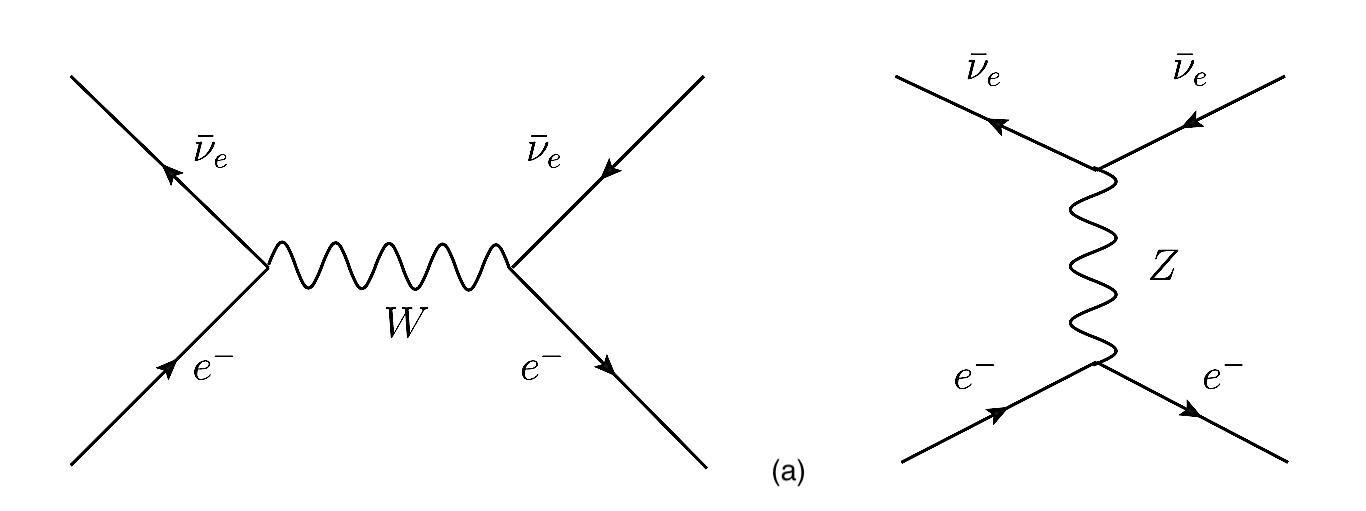}
	\includegraphics[width=3.5cm,height=4cm]{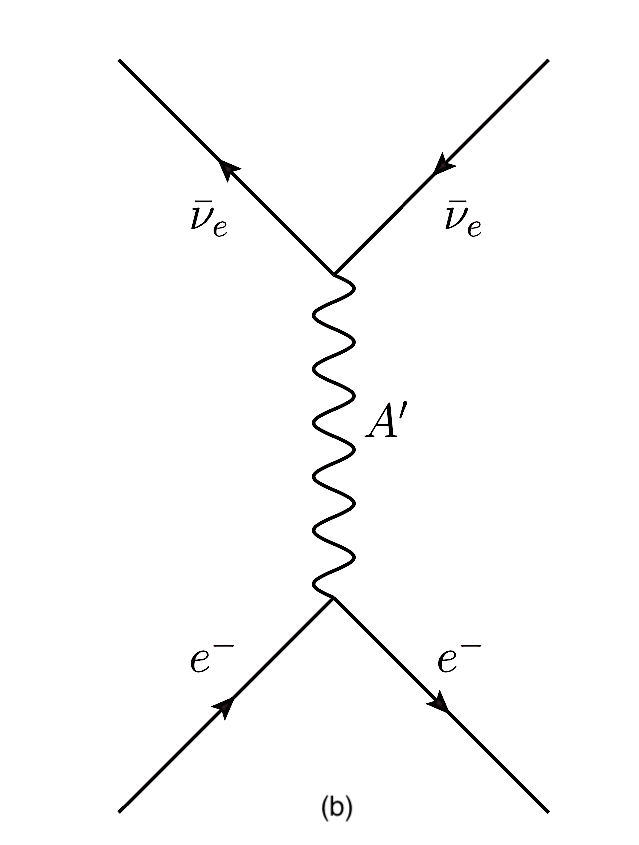}
		\caption{ \small \label{fig0} (a) Contributing Feynman diagrams for anti-electron-neutrino and electron scattering via the $W$ and $Z$ boson exchanges. Note that muon- and tau-neutrinos (including anti-muon- and anti-tau-neutrinos) scatter with electron only via the $Z$ boson exchange. (b) The Feynman diagram 
		showing the scattering of (anti-)neutrino with electron via the 
		$t$-channel dark photon $A'$ exchange.}
\end{figure}

We obtain the differential cross section for neutrino-electron scattering as a function of the recoil energy. The effective renormalizable Lagrangian given in Eq.~(\ref{Eq.3}), where the SM photon and dark photon mixed via a kinetic term. This mixing has been studied widely in literature \cite{Essig:2013lka,Essig:2010gu,Essig:2009nc}. The $U(1)_{B-L}$ gauging induces a new coupling $g_{B-L}$ in addition to its mass $M_{A'}$ and $\epsilon$. The parameter space of the current dark photon model contains a set of three variables ($M_{A'}, \epsilon, g_{B-L}$). In this study we focus on the dark model with two parameters, $M_{A'}$ and $g_{B-L}$. Figure~\ref{fig0}(b) shows the Feynman diagram of neutrino-electron scattering mediated by the dark photon.

The scattering cross section from the pure contribution of $A'$ is expressed as follows \cite{Bilmis:2015lja}.
\begin{equation}\label{Eq.6}
\big[\frac{d\sigma}{dT}(\nu e^-\rightarrow \nu e^-)\big]_{DP}=\frac{g^4_{B-L}m_e}{4\pi E^2_\nu(M^2_{\rm A'}+2m_eT)^2}(2E_\nu^2+T^2-2TE_\nu-m_eT)
\end{equation}
where the cross section is flavor blind. 
On the other hand, 
the interference term in the differential cross-section 
depends on the neutrino flavor and whether it is the
neutrino or anti-neutrino.  The interference terms are given by
\cite{Bilmis:2015lja}.,
\begin{equation}\label{Eq.7}
\frac{d\sigma_{\rm INT}(\nu_e e^-)}{dT}=\frac{g^2_{B-L}G_Fm_e}{2\sqrt{2}\pi E^2_\nu(M^2_{\rm A'}+2m_eT)}(2E_\nu^2-m_eT+\beta)
\end{equation}
\begin{equation}\label{Eq.8}
\frac{d\sigma_{\rm INT}(\bar{\nu}_e e^-)}{dT}=\frac{g^2_{B-L}G_Fm_e}{2\sqrt{2}\pi E^2_\nu(M^2_{\rm A'}+2m_eT)}(2E_\nu^2+2T^2-T(4E_\nu+m_e)+\beta)
\end{equation}
\begin{equation}\label{Eq.9}
\frac{d\sigma_{\rm INT}(\nu_\alpha e^-)}{dT}=\frac{g^2_{B-L}G_Fm_e}{2\sqrt{2}\pi E^2_\nu(M^2_{\rm A'}+2m_eT)}(-2E_\nu^2+m_eT+\beta)
\end{equation}

\begin{equation}\label{Eq.10}
\frac{d\sigma_{\rm INT}(\bar{\nu}_\alpha e^-)}{dT}=\frac{g^2_{B-L}G_Fm_e}{2\sqrt{2}\pi E^2_\nu(M^2_{\rm A'}+2m_eT)}(-2E_\nu^2-2T^2-T(4E_\nu+m_e)+\beta)
\end{equation}
where $\beta$ is given by
\begin{equation*}
    \beta=\sin^2\theta_w(8E^2_\nu-8E_\nu T-4m_e T+4T^2).
\end{equation*}
Here $\alpha$ in $\nu_\alpha$ denotes either $\mu$ or $\tau$. 
The $\nu_\mu$-$e^-$ and $\nu_\tau$-$e^-$ scatterings can 
occur through the SM $Z$ and the dark photon $A'$ mediated 
processes, while the $\nu_e$-$e^-$ scattering is mediated 
by the SM $Z,W$ bosons and dark photon $A'$. 
The pure contribution of the dark photon cross section is 
proportional to the fourth power of $g_{B-L}$ while the 
interference term is proportional to the second power of $g_{B-L}$. The total cross-section $\sigma_{T}$ is then given by $\sigma_{T}=\sigma_{DP}+\sigma_{INT}+\sigma_{SM}$ .

\section{Sensitivity reach on parameter space of the Dark Photon model at the Forward physics Experiments}
Proton-proton collisions at the Large Hadron Collider (LHC)
produce many particles along the beam axis. Forward Physics Facilities (FPF)~\cite{Feng:2022inv,Anchordoqui:2021ghd,FASER:2019aik,FASER:2021ljd,FASER:2021cpr,FASER:2020gpr,FASER:2018bac,Kling:2021gos,FASER:2019dxq,Ismail:2020yqc}, situated several hundred meters away from the ATLAS interaction point and protected by concrete and rock, will host a wide variety of experiments probing the SM processes and also searching for new physics beyond the standard model (BSM)~\cite{Jodlowski:2020vhr,Feng:2022inv,Feng:2017vli,Kling:2018wct,Feng:2018pew,Deppisch:2019kvs,Cheung:2021tmx,Bakhti:2020szu,Jho:2020jfz,Okada:2020cue,Bahraminasr:2020ssz,Kelly:2020pcy,Falkowski:2021bkq,FASER:2018eoc,Cottin:2021lzz,Ismail:2021dyp,Ansarifard:2021elw,Cheung:2022oji,Batell:2021blf,Batell:2021aja,Batell:2021snh}. 
Each experiment in the FPF is uniquely optimized for its own physics 
goals.

In this section, we study the neutrino-electron scattering mediated by the $t$-channel dark photon exchange at the running FASER$\nu$ and the proposed FASER$\nu2$, FLArE (Forward Liquid Argon Experiment), and SND$@$LHC, where $\rm FASER\nu2$ and FLArE are 
parts of the proposed Forward Physics Facility (FPF). FPF also contains other detectors called the FASER2, Advanced SND, and FORMOSA. The dark photon sensitivity at FASER2 was discussed in the literature~\cite{Feng:2022inv,Bauer:2018onh,FASER:2018eoc}.  

The FASER$\nu$ and SND@LHC detectors are made of tungsten with masses of 1.1 tons and 830 kg respectively. FASER$\nu$ is made of 1000 layers of emulsion films interleaved with tungsten plates of thickness of 1 mm. The size of FASER$\nu$ is 
$25\,{\rm cm}\times 30\,{\rm cm}\times 1.1\,{\rm m}$ \cite{talk-2022,talk-2022(2)}. That of SND@LHC is $39\,{\rm cm}\times 39\, {\rm cm}$ \cite{talk-2022(2)}. The upgraded version of FASER$\nu$ -- FASER$\nu2$ will
have a size $50\,{\rm cm}\times 50\, {\rm cm}\times 5\,{\rm m}$ 
and will weigh 10 tons. FLArE is composed of 10 ton-scale liquid argon time projection chamber (LArTPC) of the type being employed in several modern neutrino experiments. The size of the FLArE detector is $1\,{\rm m}\times1\,{\rm m}\times 7\,{\rm m}$ and will weigh 10 tons.

We are interested in the neutrino-electron scattering at the aforementioned detectors. It is well known that huge number of hadrons, such as pions, kaons and other hadrons, are produced along the beam direction. These hadrons will decay during the flight, thus producing a lot of neutrinos of all three flavors at very high energy up to a few TeV\cite{Kling:2021gos}. It was shown \cite{FASER:2020gpr} that muon neutrinos are mostly produced from charged-pion decays, electron neutrinos from hyperon, kaon and $D$-meson decays, and tau neutrinos from $D_s$ meson decays. With average energies ranging from 600 GeV to 1 TeV, the spectra of the three neutrino flavors cover a broad energy range.

The uncertainties in the coming neutrino flux were studied in
Ref.~\cite{Kling:2021gos}. 
The muon-neutrino flux has less than 10\% uncertainty up to 1 TeV,
the electron-neutrino flux has less than 10\% uncertainty up to 0.8 TeV, and 
the tau-neutrino flux has less than 10\% uncertainty up to $0.3 \sim 0.4$ TeV.
Afterwards, the uncertainties are within a factor of 2. Such uncertainties 
will propagate to the event rates predictions of order a few percent 
up to 20\%. 
Such uncertainties  are included in the systematic uncertainties.

\subsection{ Dark photon interactions at FASER$\nu$ and FASER$\nu2$ }
In this subsection we discuss the neutrino-electron $\nu e^-\rightarrow\nu e^-$ scattering at FASER$\nu$ and FASER$\nu2$ mediated by a new neutral $U(1)_{B-L}$ gauge field $A'$. The cross-section of the fixed target neutrino-electron scattering is given in Sec.~III.  We consider the dark photon of mass $M'_{A}$ ranging from $10^{-6}$ GeV to 10 GeV with an incoming neutrino beam energy $E_{\nu}$ from 10 GeV to 10 TeV. The cross section is calculated by setting the value of $g_{B-L}=10^{-2}$. We use the neutrino fluxes and energy spectra obtained in \cite{FASER:2019dxq,Kling:2021gos}
to study the neutrinos that pass through FASER$\nu$ and FASER$\nu2$. We estimate the 90\% C.L.
sensitivity reach on the parameter space of the dark photon model. We show that the best sensitivity
can be achieved in the small $M_{A'}$ region. 

\subsubsection{$\rm FASER\nu$}
The FASER$\nu$ neutrino detector is located in front of the FASER main detector~\cite{FASER:2019dxq,FASER:2018bac}, which is aligned along the collision axis to maximize the neutrino interaction. 
FASER$\nu$\cite{FASER:2019dxq} is 
a $ 25\,{\rm cm} \times 30\, {\rm cm} \times  1.1\,{\rm m}$
emulsion detector, consisting of 770 layers of emulsion films interleaved
with 1-mm-thick tungsten plates with mass 1.2 tons \cite{talk-2022} .The features of high density and shorten radiation length of tungsten are advocated as the material for the detector, because 
such features help keeping the detector small in size and 
localizing electromagnetic showers in a small volume.

\begin{figure}[h!]
	\centering
	\includegraphics[width=14cm,height=9cm]{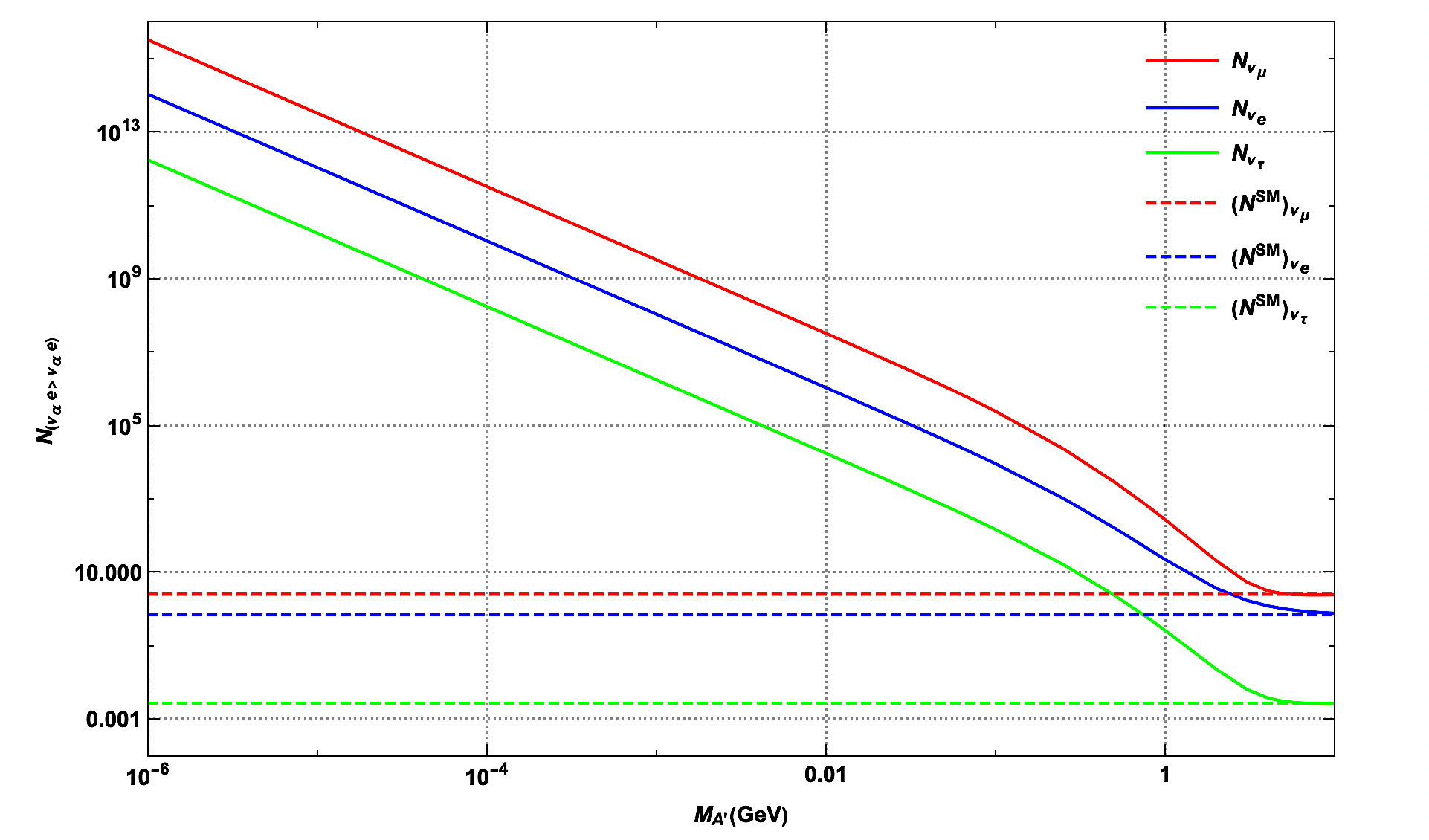}
	\caption{ \small \label{fig1} 
	Total number of active neutrinos produced at FASER$\nu$ from neutrino-electron scattering ($\nu_{\alpha} e^{-} \rightarrow \nu_{\alpha} e^{-}$), where $\alpha=e, \mu, \tau$. 
Contributions from neutrinos and anti-neutrinos are 
	included.
	}
\end{figure}

The expected numbers of neutrino events from the three flavors of active neutrinos with different dark photon masses $M_{A'}$ in FASER$\nu$ are shown in Fig. \ref{fig1}. 
The solid lines represent the number of neutrino events as a function of dark photon mass $M_{A'}$ while the dashed lines represent the SM event rates. The number of events associated with dark photon overlaps with the SM value as the dark photon mass $M_{A'}$ increases.
The highest number of events comes in with the muon neutrino $\nu_\mu$ scattering with electron($\nu_\mu e^-\rightarrow\nu_\mu e^-$), while the lowest one is from the tau neutrino $\nu_\tau$ scattering
($\nu_\tau e^-\rightarrow\nu_\tau e^-$), and that from the electron neutrino $\nu_e$ scattering is in between. Total number of events follows the same behavior as the scattering cross section. We can summarize the results in Fig. \ref{fig1} as follows, $N_{\nu_\tau}(g_{B-L},M_{A'})<N_{\nu_e}(g_{B-L},M_{A'})<N_{\nu_\mu}(g_{B-L},M_{A'})$.

To estimate the sensitivity reach in the parameter space ($g_{B-L}$ vs $M_{A'}$) of the dark photon model, we
first calculate the predicted number of events $N_{BSM}$ for the dark photon model and the SM number
of events $N_{SM}$, and treat the statistical error as $\sqrt{N_{BSM}}$ and systematic uncertainty $\sigma$
as a fraction ($\sigma$= 20\%, 5\%) of the normalization of the SM predictions. We then define
the measure of $\chi^2$ as a function of ($g_{B-L}$) and a nuisance parameter $\alpha$ as follows \cite{Cheung:2021tmx,Ballett:2019xoj}

    \begin{eqnarray}\label{Eq.11}
  \chi^2 ( g_{B-L},\alpha) &=& \min_{\alpha} \Biggr [ 
   \frac{(N^{\nu_e}_{BSM}-(1+\alpha)N^{\nu_e}_{SM})^2}{N^{\nu_e}_{BSM}} + \frac{(N^{\nu_\mu}_{BSM}-(1+\alpha)N^{\nu_\mu}_{SM})^2}{N^{\nu_\mu}_{BSM}} \nonumber \\
   &&+ \frac{(N^{\nu_\tau}_{BSM}-(1+\alpha)N^{\nu_\tau}_{SM})^2}{N^{\nu_\tau}_{BSM}}
   + \left(\frac{\alpha}{\sigma} \right)^2 \Biggr ] \;,
\end{eqnarray}
where $N_{BSM}=N_{DP}+N_{INT}+N_{SM}$ and the minimization is over the nuisance parameter $\alpha$. Here $N_{DP}$ is the number of events from the dark photon mediated process only, $N_{INT}$ is the interference term. Here we have treated the systematic uncertainties in each neutrino flavor to be the same and use
only one nuisance parameter $\alpha$. Physics-wise the systematic uncertainties come from 
theoretical calculations, the flux of neutrinos from the ATLAS IP, detector response, etc.
\begin{figure}[h!]
	\centering
	\includegraphics[width=16.8cm,height=8cm]{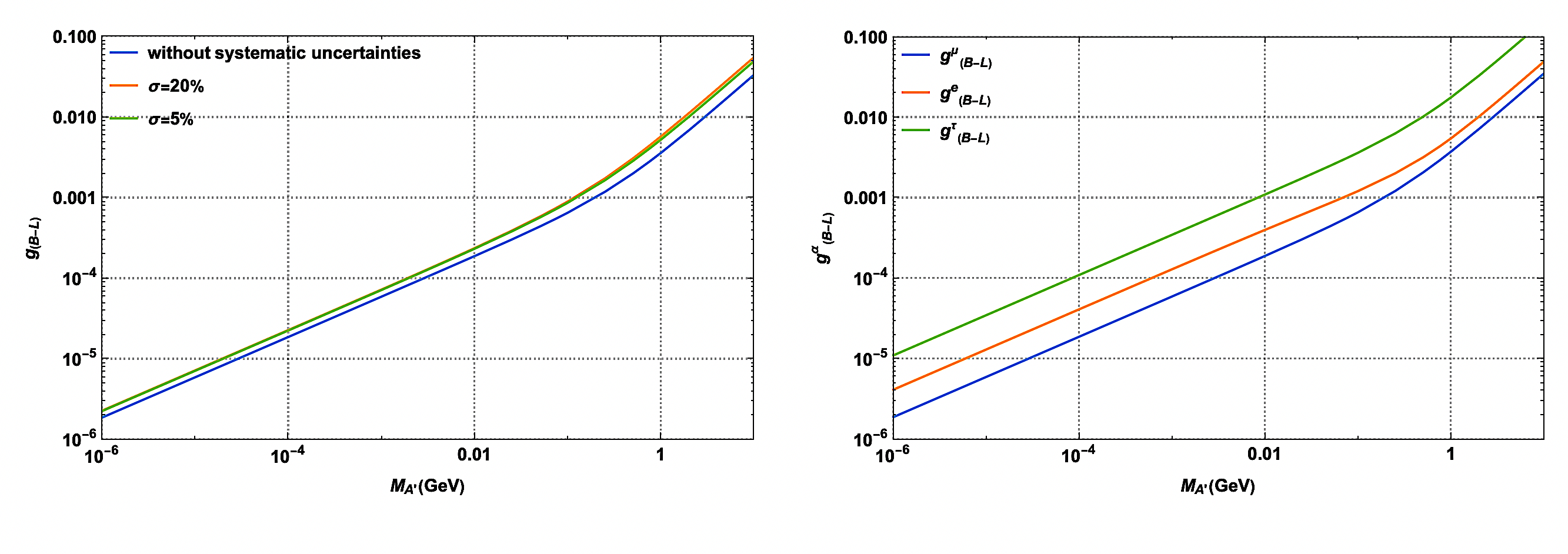}
	\caption{ \small \label{fig2}
	Left Panel: Sensitivity reach on the coupling $g_{B-L}$ versus the dark photon mass $M_{A'}$ achieved at FASER$\nu$. Systematic uncertainties $\sigma$ = 5\%, 20\% and without systematic uncertainties are shown. Right Panel: sensitivity reach on the coupling $g^{\alpha}_{B-L}$  for each neutrino flavor $\alpha=e,\mu,\tau$ versus the dark photon mass $M_{A'}$ achieved at FASER$\nu$ (without systematic uncertainties) are shown.}
\end{figure}
We show in the left panel of Fig.~\ref{fig2} the 90\% C.L. sensitivity reach (corresponding to $\chi^2 = 2.71$) 
of the coupling strength $g_{B-L}$ versus the dark photon mass $M_{A'}$ at FASER$\nu$. The higher the systematic uncertainty
the weaker the limit on $g_{B-L}$ will be. Nevertheless, the differences among
$\sigma = 5\%, 20\%$ and without systematic uncertainties  are relatively small. The sensitivity reach on
$g_{B-L}$ is the best at very small $M_{A'}$, around $2\times10^{-6}$ at $M_{A'} = 10^{-6}$ GeV 
and reduces to about $2\times10^{-2}$  at $M_{A'} = 10$ GeV.

FASER$\nu$ is primarily designed for the purpose of identifying the flavors of neutrinos \cite{FASER:2019dxq}. The expected
sensitivity for each flavor at FASER$\nu$ is shown in the right panel of Fig. \ref{fig2}. We only show the curves with no systematic uncertainty included. The curves can be compared to the corresponding one
“Without Systematic” of the left panel of Fig. \ref{fig2}. For the evaluation of $\chi^2$, we consider the special case of Eq. \ref{Eq.11} with $\alpha=0$ and it reduces to
\begin{equation}\label{Eq.12}
    \chi^2(g^{\alpha}_{B-L}) = [\frac{(N^{\nu_\alpha}_{BSM}-N^{\nu_\alpha}_{SM})^2}{N^{\nu_\alpha}_{BSM}}]
\end{equation}
where $\alpha=e,\mu,\tau$.
The green and orange curves of Fig.~\ref{fig2} depict the sensitivity reach of $g^{\tau}_{B-L}$ and 
$g^{e}_{B-L}$ ($\chi^2=2.71$) versus $M_{A'}$, respectively, while the blue curve represents the sensitivity reach of $g^{\mu}_{B-L}$ versus $M_{A'}$. It is clear from the right and left panel of Fig.~\ref{fig2} that the overall sensitivity reach of $g_{B-L}$ is dominated by $g^{\mu}_{B-L}$. The sensitivity reach by the  neutrino flavors is in the order of $\nu_\tau<\nu_e<\nu_\mu$.  

\subsubsection{FASER$\nu2$}
FASER$\nu2$ is a much more advanced version of FASER$\nu$\cite{Anchordoqui:2021ghd}. FASER$\nu2$ detector is currently proposed to be composed of 3330 emulsion layers interleaved with 2~mm thick tungsten plates. It will also accommodate a veto detector and an interface detector to the FASER2 spectrometer, with one detector in the middle of the emulsion modules and the other detector downstream of the emulsion modules to make the global analysis and muon charge measurement possible.
Both the emulsion modules and interface detectors will be put in a cooling system. The total volume of the tungsten target is 
$50\,{\rm  cm} \times 50\, {\rm cm} \times 5\, {\rm m}$, and the mass is 10 tons. The detector length, including the emulsion films and interface detectors, will be about 8~m.

\begin{figure}[h!]
	\centering
	\includegraphics[width=14cm,height=9cm]{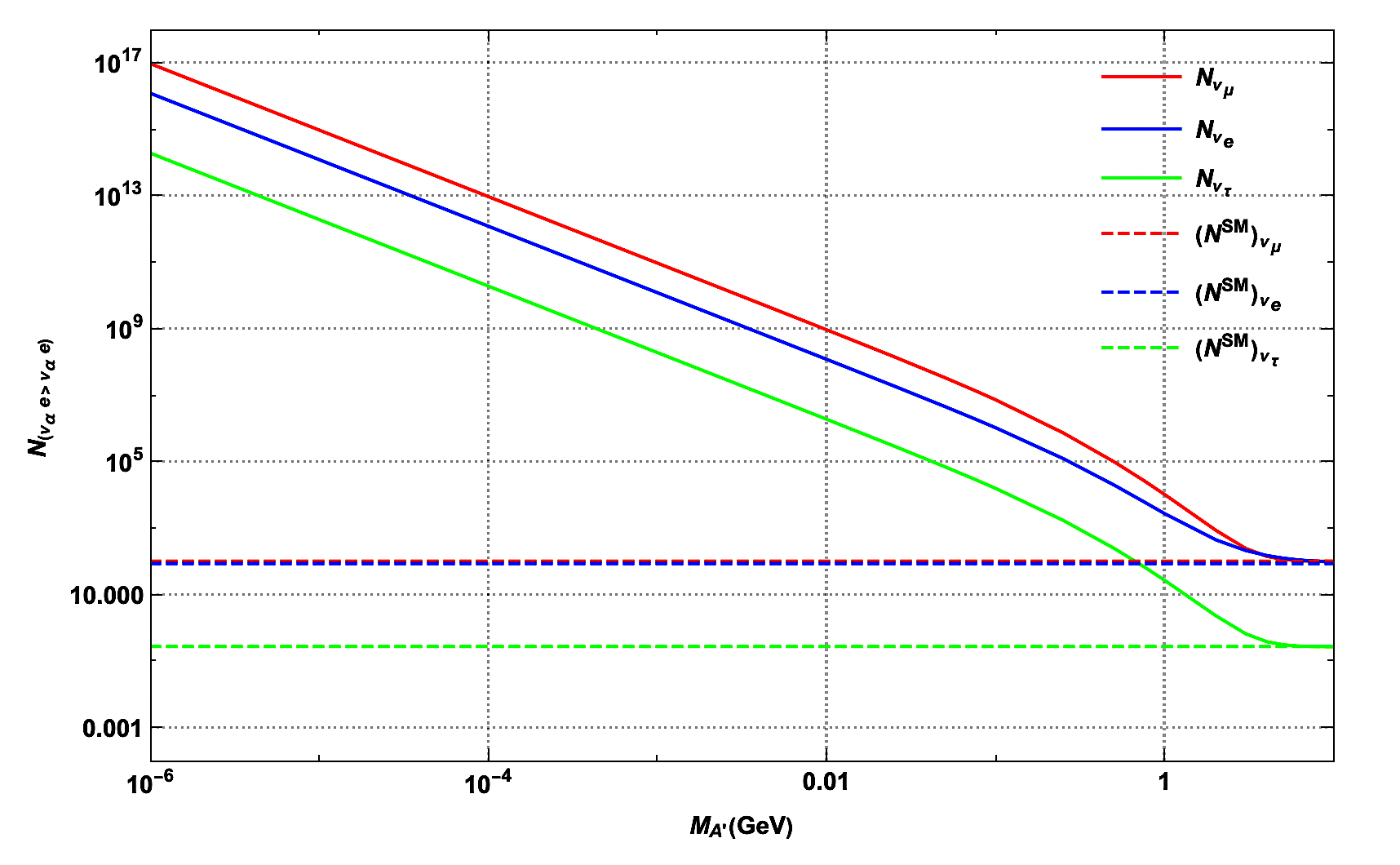}
	\caption{ \small \label{fig3}
	Total number of active neutrinos produced at FASER$\nu2$ from neutrino-electron scattering ($\nu_{\alpha} e^{-} \rightarrow \nu_{\alpha} e^{-}$), where $\alpha=e, \mu, \tau$. Contributions from neutrinos and anti-neutrinos are 
	included.
	}
\end{figure}

 In this subsection we calculate the sensitivity reach on the parameter space of the dark photon model at FASER$\nu2$. 
 We follow the same procedures as the last subsection for the estimation of the total number of neutrino events at FASER$\nu2$. The total numbers of neutrino events at FASER$\nu2$ are shown 
 in Fig. \ref{fig3}, which follows the same notation as 
 Fig.~\ref{fig1}.
Comparing with FASER$\nu$, FASER$\nu2$ could collect a larger number of neutrino events. There are sizable increments in 
both the BSM neutrino events and SM neutrino events. In the heavy $M_{A'}$ mass regime ($M_{A'} \agt 5$ GeV),
the $N_{\nu_\mu}$ curve overlaps with the $N_{\nu_e}$ curve. 
This behavior arises from the large interference effect 
associated with the muon neutrinos. Among the three neutrino 
flavors, the tau neutrino is the least in event rates.

\begin{figure}[th!]
	\centering
	\includegraphics[width=16.8cm,height=8cm]{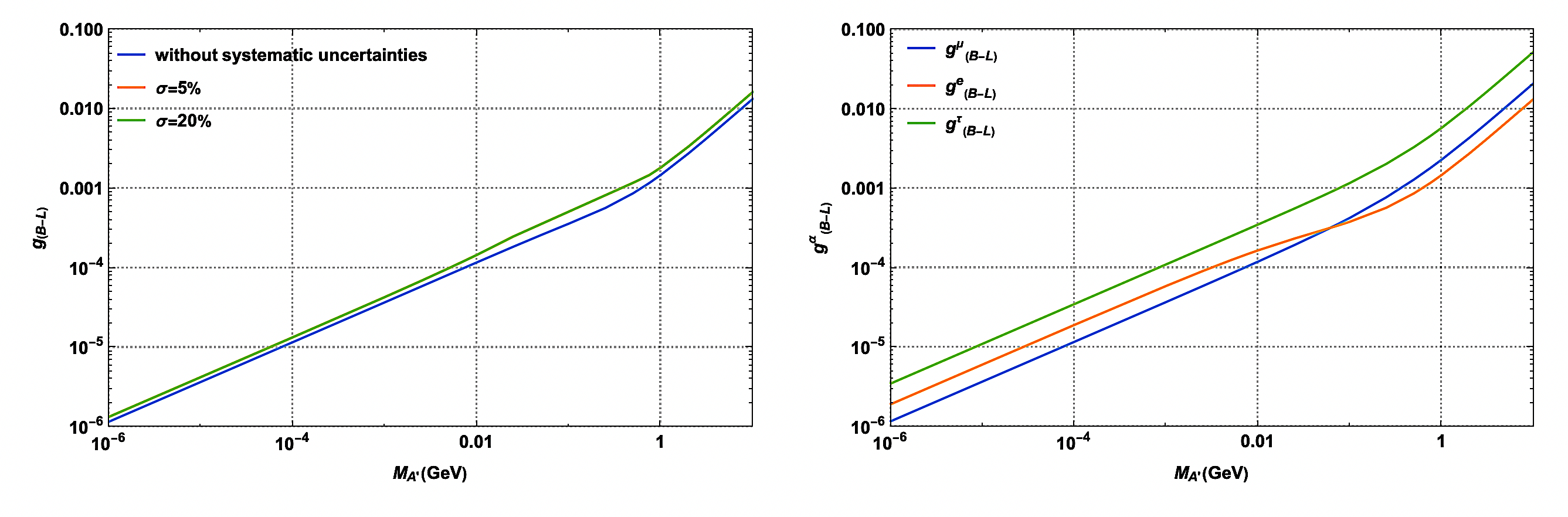}
	\caption{ \small \label{fig4}
	Left Panel: Sensitivity reach on the coupling $g_{B-L}$ versus the dark photon mass $M_{A'}$ achieved at FASER$\nu2$. Systematic uncertainties $\sigma$ = 5\%, 20\% and without systematic uncertainties are shown. Right Panel: sensitivity reach on the coupling $g^{\alpha}_{B-L}$ for each neutrino flavor $\alpha=e,\mu,\tau$ versus the dark photon mass $M_{A'}$ achieved at FASER$\nu2$ (without systematic uncertainties) are shown.
	}
\end{figure}

Sensitivity reach at FASER$\nu2$ is carried out by a similar $\chi^2$ analysis in Eq.~(\ref{Eq.11}). We estimate the 90\% C.L. sensitivity reach (corresponding to $\chi^2 = 2.71$) 
of the coupling strength $g_{B-L}$ versus dark photon mass 
$M_{A'}$ at FASER$\nu2$, as shown in Fig.~\ref{fig4}.
The dark photon sensitivity reach at FASER$\nu2$ is shown in the left panel of Fig. \ref{fig4}, in which the curves with $\sigma=5\%$ and and $20\%$ are overlapping with each other.
FASER$\nu2$ has better sensitivity compared to FASER$\nu$.
In the very small dark photon mass region, $M_{A'}=10^{-6}\rm ~GeV$, FASER$\nu2$ can reach as low as $10^{-6}$ while
FASER$\nu$ only reaches around $2\times10^{-6}$.
The sensitivity devalues with the increment of dark photon mass.

The expected sensitivity reach of FASER$\nu 2$ for each neutrino flavor is given in the right panel of Fig.~\ref{fig4}. The effect of interference is clearly visible from the cross-over of $g^{e}_{B-L}$ curve over the $g^{\mu}_{B-L}$ curve. In the higher $M_{A'}$ mass region, the $g^e_{B-L}$ curve has better sensitivity compared with the $g^{\mu}_{B-L}$ curve. The total sensitivity 
curve of $g_{B-L}$ in the left panel of Fig.~\ref{fig4} resembles the $g^{\mu}_{B-L}$ curve of the right panel in the low $M_{A'}$ region and switches to $g^{e}_{B-L}$ in the heavier $M_{A'}$ region.

\subsection{ Dark photon interactions at SND@LHC }

\begin{figure}[h!]
	\centering
	\includegraphics[width=14cm,height=9cm]{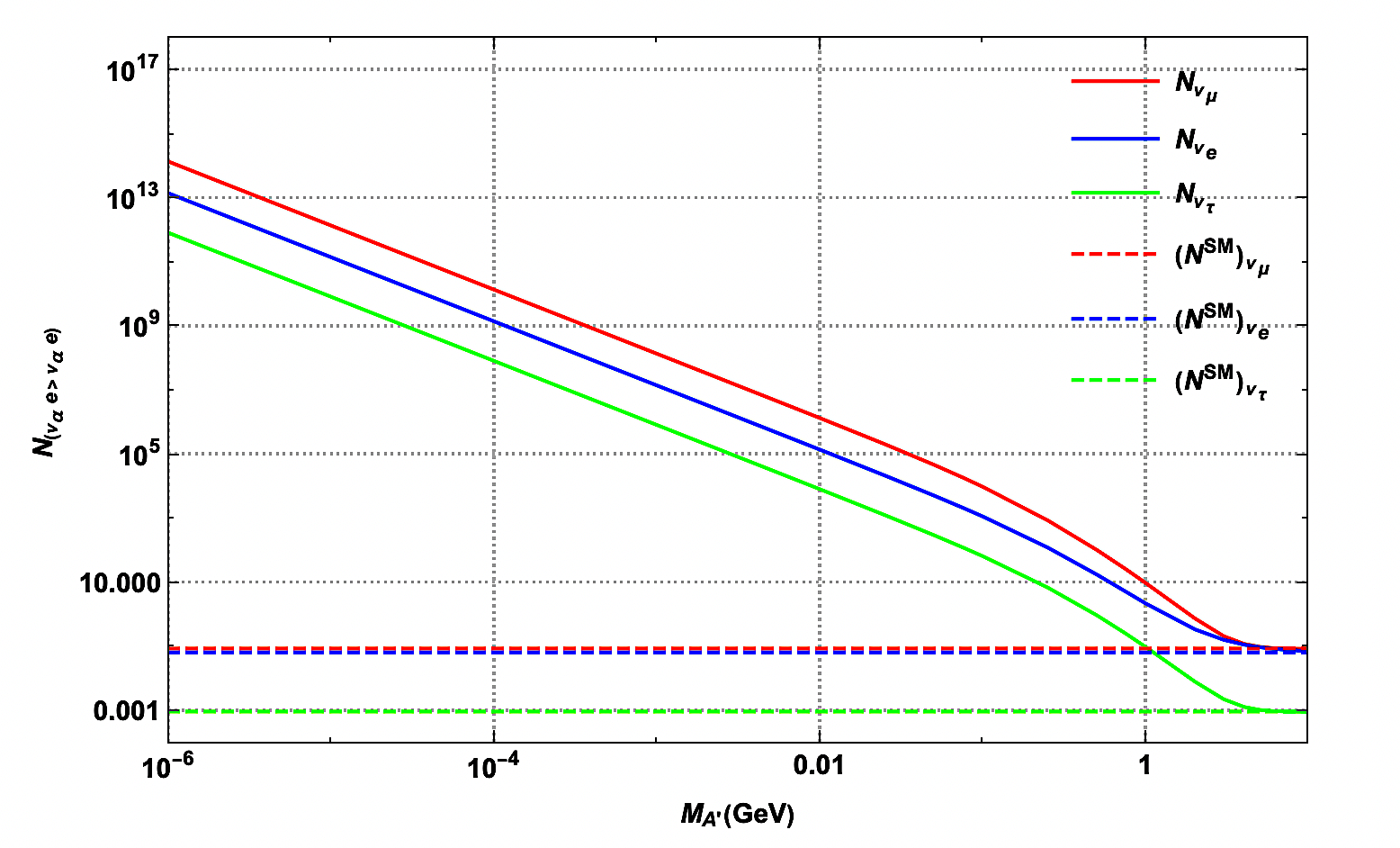}
	\caption{ \small \label{fig5}
	Total number of active neutrinos produced at SND@LHC from neutrino-electron scattering ($\nu_{\alpha} e^{-} \rightarrow \nu_{\alpha} e^{-}$), where $\alpha=e, \mu, \tau$.
	Contributions from neutrinos and anti-neutrinos are 
	included.
	}
\end{figure}

SND@LHC is a compact detector of neutrinos at the LHC and 
it consists of a neutrino target followed downstream by a device for detecting muons, which are produced when neutrinos interact with the target.  It is especially complementary to the FASER$\nu$. %
Targets are made from tungsten plates interlaced with emulsion films and electronic tracking devices. Emulsion films display the tracks of particles produced in neutrino interactions, while electronic tracking devices provide the time stamps for these tracks. The tracking devices also measure the energy of the neutrinos along with the muon detector.

The predicted number of neutrino events mediated by the 
dark photon of mass ranging from $M_{A'} \sim 
10^{-6}-10 \rm ~ GeV$ is shown in Fig. \ref{fig5}. 
Comparing with FASER$\nu$ and FASER$\nu2$, SND@LHC has fewer neutrino events.  We summarize the predicted numbers of neutrino events for these three forward-physics experiments as follows, 
$\rm N^{\rm SND@LHC}_{BSM}(g_{B-L},M_{A'})<N^{FASER\nu}_{BSM}(g_{B-L},M_{A'})<N^{FASER\nu2}_{BSM}(g_{B-L},M_{A'})$. The difference 
in the event rate arises from the neutrino flux at each detector volume.

\begin{figure}[h!]
	\centering
	\includegraphics[width=16.8cm,height=8cm]{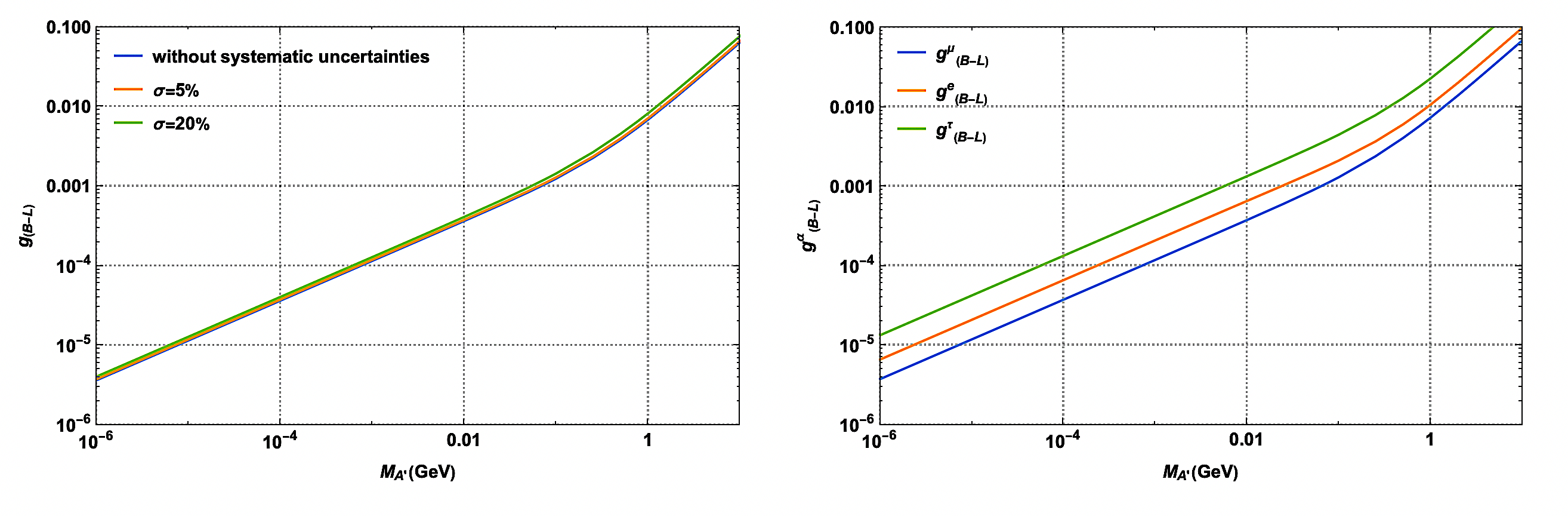}
	\caption{ \small \label{fig6}
	Left Panel: Sensitivity reach on the coupling $g_{B-L}$ versus the dark photon mass $M_{A'}$ achieved at SND@LHC. Systematic uncertainties $\sigma$ = 5\%, 20\% and without systematic uncertainties are shown. Right Panel: sensitivity reach on the coupling $g^{\alpha}_{B-L}$  for each neutrino flavor $\alpha=e,\mu,\tau$ versus the dark photon mass $M_{A'}$ achieved at SND@LHC (without systematic uncertainties) are shown.
	}
\end{figure}

A similar $\chi^2$ analysis (Eq. \ref{Eq.11}) is carried out at SND@LHC to assess dark photon sensitivity.  Accordingly, a 90\% C.L. sensitivity reach is estimated (corresponding to $\chi^2 = 2.71$) for the coupling strength $g_{B-L}$ versus the dark photon mass $M_{A'}$ at the SND@LHC, as shown in 
the left panel of Fig. \ref{fig6}. Here we
have assumed a benchmark detector made of tungsten with dimensions $39\,{\rm  cm}\times39\, {\rm cm}$
at the 14 TeV LHC with an integrated luminosity of $L =150~fb^{-1}$. The higher the systematic uncertainty
the weaker the limit on $g_{B-L}$ will be. Nevertheless, the differences among
$\sigma = 5\%, 20\%$ and without systematic uncertainties  are relatively small. The sensitivity reach on
$g_{B-L}$ is the best at very small $M_{A'}$ as low as  $4\times10^{-6}$ at $M_{A'} = 10^{-6}$ GeV 
and devalues to about $8\times10^{-2}$  at $M_{A'} = 10$ GeV. 
The reach of $g^{\alpha}_{B-L}$  for each neutrino flavor as 
a function of dark photon mass $M_{A'}$ is shown in the 
right panel of Fig. \ref{fig6}.
\begin{figure}[h!]
	\centering
	\includegraphics[width=14cm,height=9cm]{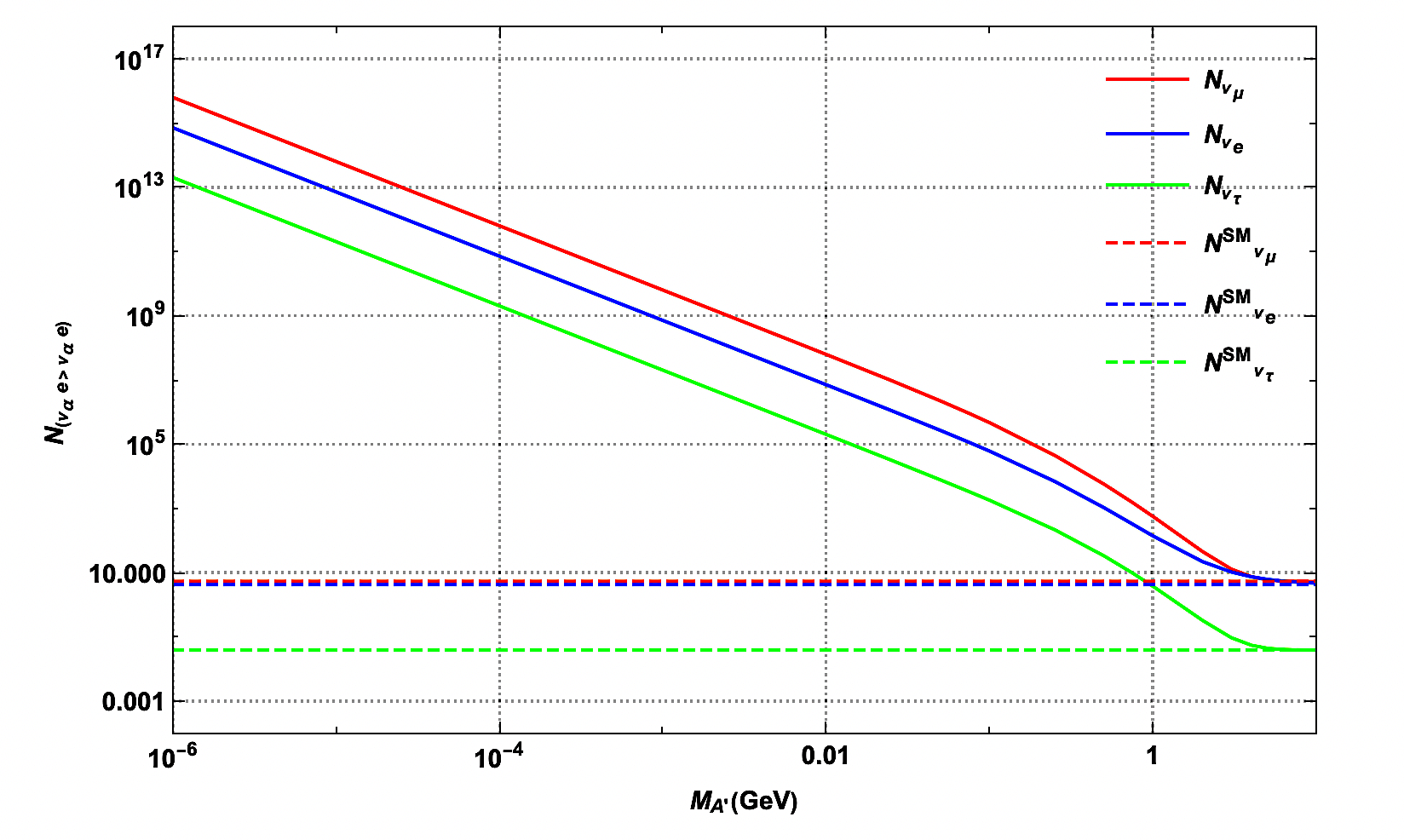}
	\caption{ \small \label{fig7}
	Total number of active neutrinos produced at FLArE (10 tons) from neutrino-electron scattering ($\nu_{\alpha} e^{-} \rightarrow \nu_{\alpha} e^{-}$), where $\alpha=e, \mu, \tau$.
Contributions from neutrinos and anti-neutrinos are 
	included.
	}
\end{figure}

\subsection{ Dark photon interactions at FLArE (10 tons)}

As a part of the suite of detectors for the FPF, the liquid argon time projection chamber (LArTPC) is being considered. This type of detector offers the capability to determine particle identification, track angle, and kinetic energy over a wide range of energies. Dark matter (DM) direct detection searches~\cite{Batell:2021blf} and neutrino experiments have successfully used liquid argon as an active detector. Detecting neutrino energy~\cite{Anchordoqui:2021ghd} and searching for dark matter are therefore well suited to be put together to 
utilize such a liquid-argon detector.
There is an imperative motivation to detect and measure TeV-scale neutrino events from a laboratory-generated and well-characterized source. Moreover, the LHC could be the only source of high-energy intense tau neutrinos on Earth.
Here we have assumed a benchmark detector made of tungsten with dimensions $1 {\rm m}\times 1\,{\rm m} \times 8\, {\rm m}$
at the 14 TeV LHC with an integrated luminosity of $L =3000~fb^{-1}$.

The expected number of neutrino events at FLArE(10 tons) is presented in Fig. \ref{fig7}. Comparing with the aforementioned forward physics experiments, 
the expected number of neutrino events at the FLArE 
as a function of dark photon mass $M_{A'}$ and coupling $g_{B-L}$
are in the following ordering:
$\rm N^{\rm SND@LHC}_{BSM}(g_{B-L},M_{A'})<N^{FASER\nu}_{BSM}(g_{B-L},M_{A'})<N^{FLArE}_{BSM}(g_{B-L},M_{A'})<N^{FASER\nu2}_{BSM}(g_{B-L},M_{A'})$. This ordering will reflect the bound on the dark photon coupling.

\begin{figure}[h!]
	\centering
	\includegraphics[width=16.8cm,height=8cm]{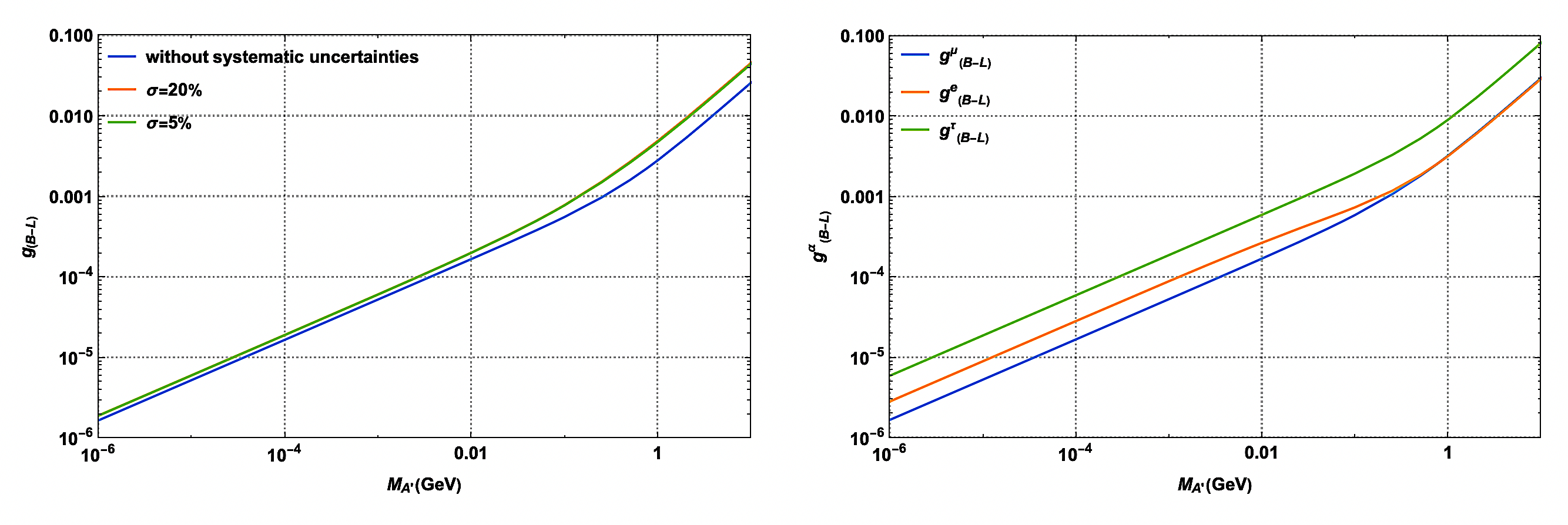}
	\caption{ \small \label{fig8}Left Panel: Sensitivity reach on the coupling $g_{B-L}$ versus the dark photon mass $M_{A'}$ achieved at FLArE (10 tons). Systematic uncertainties $\sigma$ = 5\%, 20\% and without systematic uncertainties are shown. Right Panel: sensitivity reach on the coupling $g^{\alpha}_{B-L}$  for each neutrino flavor $\alpha=e,\mu,\tau$ versus the dark photon mass $M_{A'}$ achieved at FLArE(10 tons) (without systematic uncertainties) are shown.}. 
\end{figure}

The bound on the dark-photon coupling is obtained through the $\chi^2$ defined in Eq.~(\ref{Eq.11}). We obtain the 90\% C.L. dark photon sensitivity at FLArE (10 tons), as shown in  
the left panel of the Fig.~\ref{fig8}, which shows the coupling $g_{B-L}$ versus the dark photon mass $M_{A'}$ 
with systematic uncertainties $\sigma$ = 5\%, 20\% and without systematic uncertainties. 
The right panel of Fig.~\ref{fig8} obtained by Eq.~(\ref{Eq.12})
shows the sensitivity reach on the coupling $g^{\alpha}_{B-L}$  for each neutrino flavor $\alpha=e,\mu,\tau$ versus the dark photon mass $M_{A'}$ achieved at FLArE (10 tons) (without systematic uncertainties). 
The effect of interference is visible from the 
right panel of the Fig. \ref{fig8}, in which
the $g^e_{B-L}$ curve overlaps with the $g^\mu_{B-L}$ curve 
for $M_{A'} \agt 0.1$ GeV. 

\section{Complementarity of FASER$\nu/2\nu$, SND@LHC and FLA$\textbf{r}$E Results}

 We have obtained the future sensitivity reach in the parameter space ($g_{B-L}, M_{A'}$) of the dark photon model at FASER$\nu$, FASER$2\nu$, SND@LHC, and FLArE(10 tons). In this section, we compare these forward physics experiments with existing neutrino-electron scattering experiments. While comparing the Forward Physics Experiments to one another, FASER$\nu2$ has the best sensitivity among the four experiments while  SND@LHC is the least sensitive one. 
 For the case of a very small dark photon mass $M_{A'}=10^{-6}\rm~GeV$ 
 the sensitivity reach of $g_{B-L}$ is around $4\times10^{-6}$ at SND@LHC, $2\times10^{-6}$ at FASER$\nu$, $1.6\times10^{-6}$ at FLArE(10 tons), and $10^{-6}$ at FASER$\nu2$. FASER$\nu 2$ has the best sensitivity for the
 whole range of dark photon mass considered in this work.

\begin{figure}[h!]
	\centering
	\includegraphics[width=17cm,height=10cm]{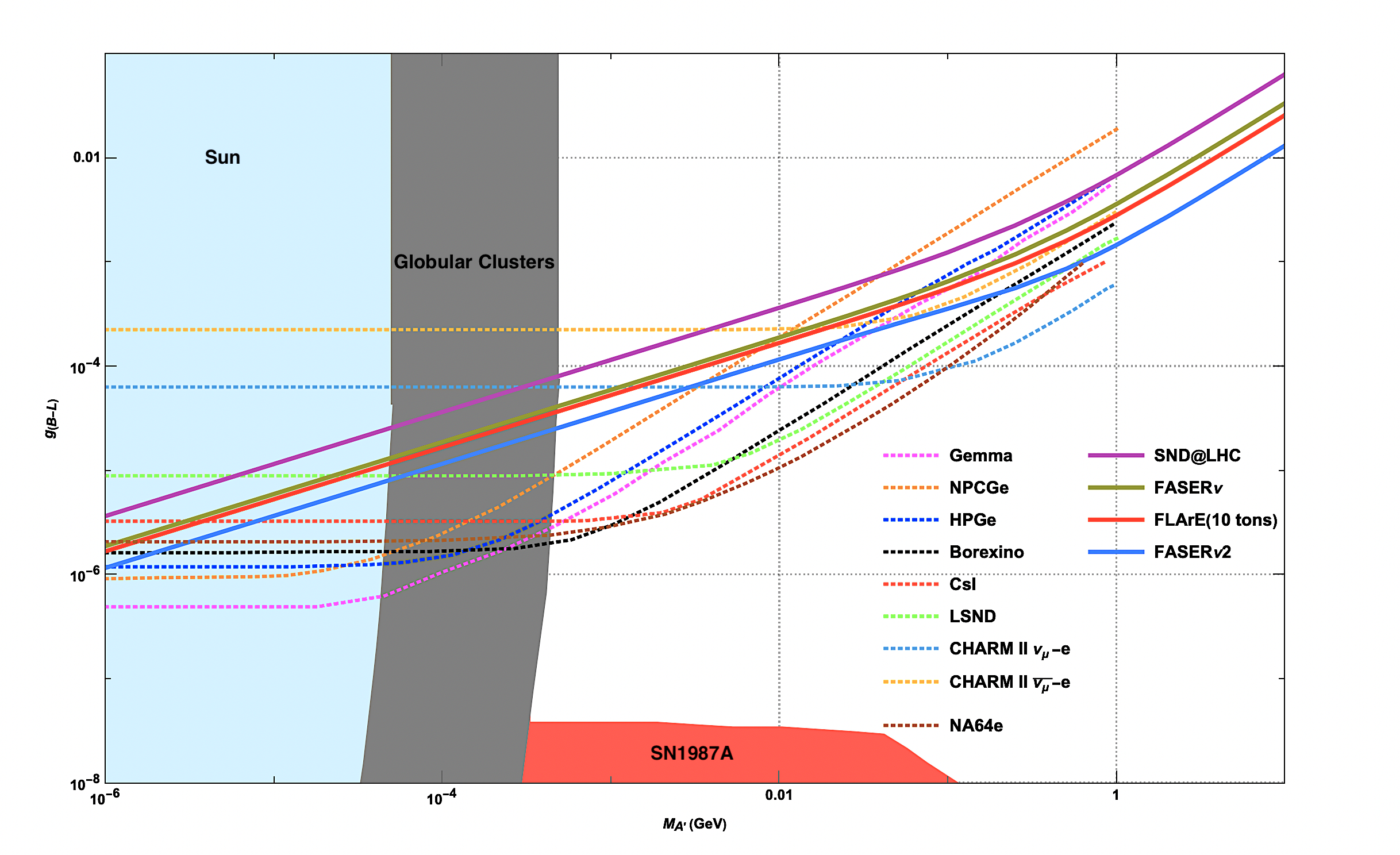}
	\caption{ \small \label{fig9}
	Future sensitivity reach of FASER$\nu$ (military green), FASER$2\nu$ (blue), SND@LHC (violet), and FLArE (10 tons)(red) at 90\% C.L.(without systematic uncertainties) on the gauge coupling $g_{B-L}$ of the $U_{B-L}(1)$ group as the function of dark photon mass $M_{A'}$. 
	Other existing constraints at 90\% C.L. shown include (i) $\bar{\nu}_e-e$ scattering measurement at Gemma (dashed magenta curve), (ii) NPCGe (dashed orange curve), (iii) HPGe (dashed blue curve), (iv) elastic scattering of neutrinos using a liquid scintillator at Borexino (dashed black curve), (v) CsI (dashed red curve) measurement of $\bar{\nu}_e-e^{-}$, (vi) LSND (dashed light green curve) measurement of $\nu_e-e^{-}$, (vii) CHARM-II-$\nu_\mu-e$ (dashed light blue), (viii) CHARM-II-$\bar{\nu}_\mu-e$ (dashed light yellow), and (ix) NA64 experiment at the CERN SPS(unbroken $U(1)_{\rm B-L}$)(dashed brown). 
 The light blue region indicates the solar constraints (luminosity analysis in the conversion of 
 plasmons in the Sun), the gray region represents the constraint
 derived from energy loss due to dark photons in globular clusters,
 and the orange region indicates the constraint due to 
 supernova cooling. These colored regions are taken from 
Ref.~\cite{Bilmis:2015lja,Harnik:2012ni}.
}
\end{figure}

The GEMMA Collaboration~\cite{Beda:2009kx} measured the $\bar{\nu}_e-e$ scattering cross section, which can be used to put a bound on neutrino magnetic moments. The Gemma (dashed magenta curve) in Fig.~\ref{fig9} 
has the best sensitivity among all the considered forward Physics experiments at the low dark photon mass region. The Gemma sensitivity gets weakened with increment of dark photon mass, after $M_{A'}>0.01~\rm GeV$ there is a considerable reduction in the Gemma sensitivity to the dark photon coupling. In the heavier mass region the forward physics experiments
estimated in this work in general have better sensitivity.

The TEXONO-NPCGe\cite{Chen:2014dsa} estimated the constraints on millicharged neutrinos via analysis of data from atomic ionizations. 
NPCGe (dashed orange curve) also has better sensitivity in the low dark photon mass region, and the sensitivity gets weakened with increases in 
dark photon mass. The sensitivity curve of NPCGe follows the similar pattern of Gemma. 
The TEXONO-HPGe~\cite{TEXONO:2002pra} estimated the limit on the 
electron-neutrino magnetic moment. FASER$\nu2$ (blue curve) has better sensitivity 
than that of HPGe (dashed blue) in the very small dark photon mass $M_{A'}\sim 10^{-6}~\rm GeV$. However, in the intermediate mass region $\rm 10^{-6 }~GeV<M_{A'}<0.01~GeV$, HPGe has better sensitivity than all the forward physics experiments considered in this work.

The Borexino Collaboration \cite{Bellini:2011rx} measured the spectrum of the  $\rm {}^{7}Be$ solar neutrino (with 862 keV energy) via the elastic scattering of neutrinos using a liquid scintillator. The sensitivity curve of Borexino (dashed black curve) also followed the similar pattern of TEXONO. However, both FLArE (red curve) and FASER$\nu2$ (blue) have better sensitivity at $M_{A'}=10^{-6}~\rm GeV$ compared with Borexino.

\begin{figure}[h!]
	\centering
	\includegraphics[width=17cm,height=10cm]{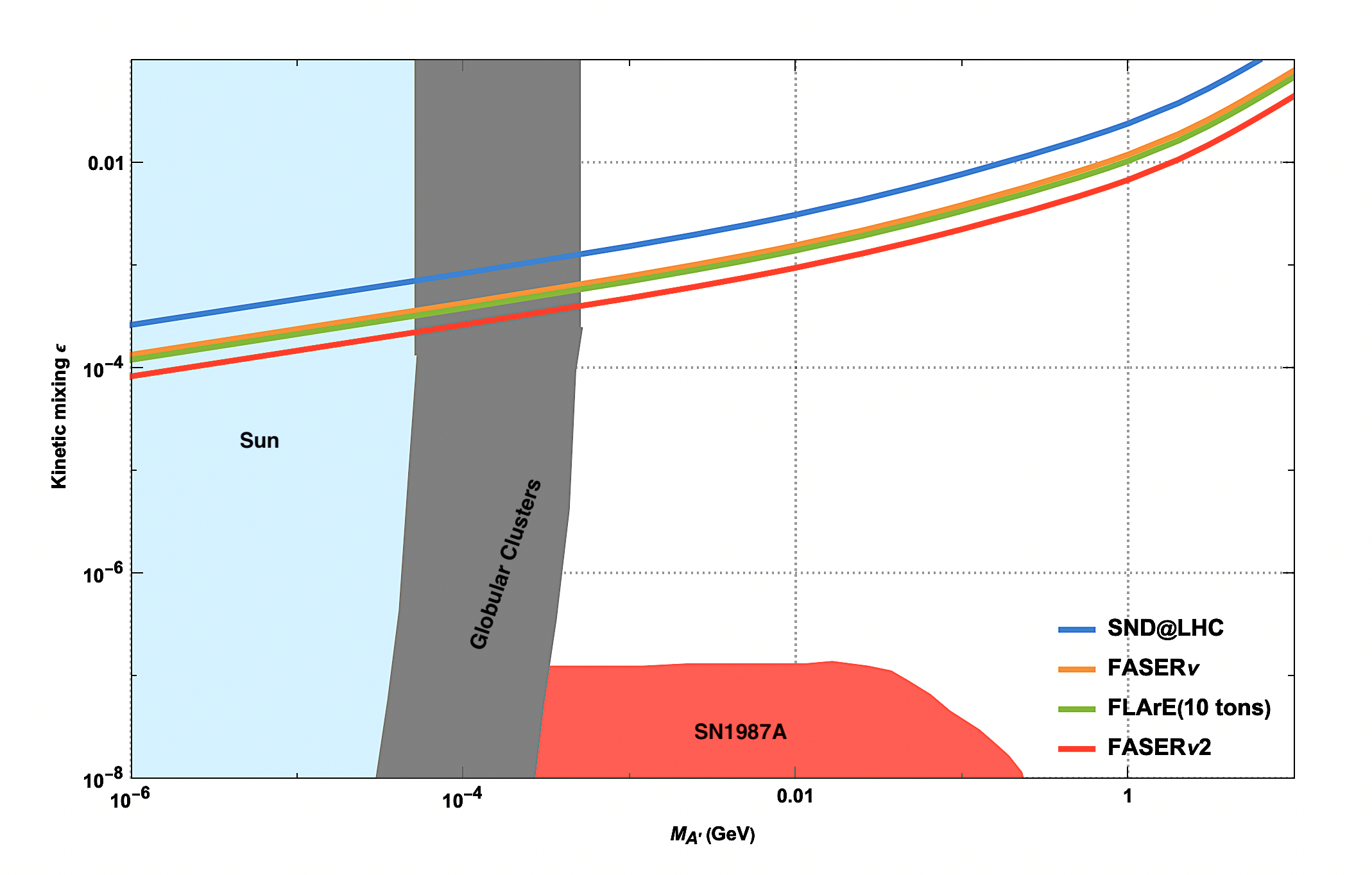}
	\caption{ \small \label{fig10}
	Future sensitivity reach of FASER$\nu$ (orange), FASER$2\nu$ (red), SND@LHC (blue), and FLArE (10 tons)(green) at 90\% C.L.(without systematic uncertainties) on the kinetic mixing parameter 
	$\epsilon$ as a function of dark photon mass $M_{A'}$. 
   The colored regions are the same as in Fig.~\ref{fig9}
   \cite{Bilmis:2015lja,Harnik:2012ni}
 	}
\end{figure}

TEXONO-CsI~\cite{TEXONO:2009knm} measured the $\bar{\nu}_e-e^{-}$ scattering. FASER$\nu$ (dark green), FASER$\nu2$ (blue), and FLArE (red curve) all have better sensitivity in the very low $M_{A'}$ mass region ($M_{A'}<2\times10^{-6}~\rm GeV$). 
However, in the heavier dark photon mass region ($M_{A'}<1\rm~GeV$) CsI has better sensitivity than any of the forward physics experiments. 
LSND~\cite{LSND:2001akn} also followed the similar pattern as CsI. However SND@LHC(violet curve) has better sensitivity compared with LSND (dashed light green) in the low $M_{A'}$ mass region. In all the above, except CsI, the forward physics experiments have better sensitivity in the heavy dark photon mass region. CHARM II~\cite{CHARM-II:1993phx} also measured the neutrino-electron scattering in the lower dark photon mass region. The forward physics experiments have better sensitivity compared to CHARM II (CHARM-II $\nu_\mu-e$ (dashed light blue) and CHARM-II $\bar{\nu}_\mu-e$ (dashed light yellow)).  A search for a new $Z'$ gauge boson associated with (un)broken $B-L$ symmetry in the keV- GeV mass range is carried out for the first time using the missing-energy technique in the NA64 experiment (dashed brown- unbroken $U(1)_{B-L}$) at the CERN SPS\cite{Andreev:2022hxz}. In the Low dark photon mass $M_{A'}\sim 10^{-6}$, the FPF experiments (exclude SND@LHC) have better sensitivity compared to NA64e.

The bounds on the kinetic mixing parameter $\epsilon$ as a function of dark photon mass $M_{A'}$ are shown in Fig. \ref{fig10}. 
%
%
The sensitivity reach on the dark photon kinetic mixing $\epsilon$ follows the same pattern as $g_{B-L}$ coupling. While comparing the bounds on $\epsilon$ achieved in the forward physics experiments with existing neutrino-electron scattering experiments, the forward physics experiments have better bounds in the very low dark photon mass region $(M_{A'}\sim 10^{-6}\rm~GeV)$ and in the heavier dark photon mass region $M_{A'}>0.1 \rm~GeV$. 
For the case of a very small dark photon mass $M_{A'}=10^{-6}\rm~GeV$ the sensitivity reach of $\epsilon$ is around $2\times10^{-4}$ for SND@LHC, 
$1.3\times10^{-4}$ for FASER$\nu$, $1.2\times10^{-4}$ for FLArE(10 tons), and $8\times10^{-5}$ for FASER$\nu2$. FASER$\nu 2$ has the best sensitivity for all dark photon mass range considered in this work. The sensitivity on $\epsilon$ gets weakened with increment of dark photon mass.

Sensitivities of the coupling $g_{B-L}$ or the kinetic mixing $\epsilon$ of the dark photon have been evaluated for the FASER and/or FASER2, as we have mentioned before.  Here we make comparisons with those in literature \cite{Feng:2022inv,Bauer:2018onh,FASER:2018eoc}. 
Note that  the sensitivities obtained at FASER/FASER2 are based on the dark photon(s) produced mainly from the decays of hadrons, such as $\pi, \eta, K$, while those obtained in this work are based on the neutrino-electron scattering in FASER$\nu$. In Ref.~\cite{Feng:2022inv,Bauer:2018onh}, sensitivities of the coupling $g_{B-L}$ at FASER (FASER2) can be down to $3\times 10^{-6} - 0.7\times 10^{-6}$  ($5 \times 10^{-7} - 4 \times 10^{-8}$)
for $M_{A'} = 2 \times 10^{-3} - 0.1$ GeV ($M_{A'} = 2 \times 10^{-3} - 1$ GeV).
In Ref.~\cite{FASER:2018eoc}, $g_{B-L}$ can be probed down to  $8 \times 10^{-7} - 2 \times 10^{-7}$
for $M_{A'} = 3\times 10^{-6} - 0.3$ GeV.  Another model Little String Theory was
studied in Ref.~\cite{Anchordoqui:2022kuw}
that the coupling $g_X$ can be probed down to
almost $10^{-8}$ for $M_X = 0.2 - 1.3$ GeV.  It is clear that the study in this work showed that the sensitivities are not as good as  those listed above.
Nevertheless, the FPF experiments studied here showed advantage for $M_{A'} > 1$ GeV.

Note that for $M_{A'} = 10^{-6} - 10^{-1}$ GeV there are more 
stringent constraints from stella and supernova cooling \cite{Dent:2012mx,Kazanas:2014mca}. We add the constraint due to coupling in 
the Sun\cite{Redondo:2008aa,Raffelt:1988rx,An:2013yfc} (light blue), globular clusters\cite{Redondo:2008aa,Raffelt:1988rx,An:2013yfc,Jaeckel:2010ni} (gray), and SN1987A\cite{Dent:2012mx,Kazanas:2014mca} (orange)
to the summary figures in Fig.~\ref{fig9} and \ref{fig10} for
$g_{B-L}$ and $\epsilon$, respectively.
For $M_{A'} = 10^{-6} - 3 \times 10^{-4}$ GeV, the whole range of
$g_{B-L},\, \epsilon = 10^{-8} - 0.1$ is excluded.  
The SN1987A constraint ruled out $g_{B-L} < 0.5 \times 10^{-7}$ 
and $\epsilon < 10^{-7}$ for $M_{A'}= 10^{-4} - 10^{-1}$ GeV.

\section{Conclusions}
Forward Physics Facilities (FPF) can host a number of experiments 
that make use of the unique opportunities of neutrino beams in the energy range of a few hundred GeV to TeV in exploring physics beyond the SM.
We have investigated sensitivity reach on a couple of 
dark photon models at a number of experiments, including
FASER$\nu$, FASER$\nu2$, SND@LHC, and FLArE(10 tons).
We employed two models: (i) $U(1)_{B-L}$ and (ii) a hidden $U(1)'$
that mixes with the SM gauge fields via the kinetic mixing (vector portal).
The presence of such a $U(1)_{B-L}$ gauge field or a dark photon
gives rise to a detectable signal in neutrino-electron scattering.
We have studied neutrino-electron scattering mediated via $t$-channel exchange of the $U(1)_{B-L}$ gauge field at the FPF experiments 
such as FASER$\nu$, FASER$\nu2$, SND@LHC and FLArE (10 tons), and 
calculated the expected sensitivity reach on the 
parameter space of the $U(1)_{B-L}$ model. 
We investigated the advantage of FPF in a wide mass range of 
the $U(1)_{B-L}$ gauge field and to determine the flavor 
dependence of the coupling between neutrino and this new gauge boson.
We found that both FASER$\nu$ and SND@LHC are more sensitive 
to $g_{B-L}^{\mu}$. On the other hand, FASER$\nu2$ is more sensitive to $g_{B-L}^{\mu}$ in the lower dark photon mass region, but shifts to 
$g_{B-L}^{e}$ in the heavier dark photon mass region 
due to the interference effect. 
A similar pattern happens in FLArE, however, in the heavier dark 
photon mass region both $g_{B-L}^{\mu}$ and $g_{B-L}^{e}$ receive
similar sensitivity.

On the other hand, the sensitivities on the kinetic mixing 
parameter $\epsilon$ are calculated 
as a function of dark photon mass $M_{A'}$.  
The bound of kinetic mixing also follows the same pattern of the gauge coupling $g_{B-L}$.

We have covered a wide mass range of dark photon $10^{-6}~\rm GeV\leq M_{A'}\leq10~GeV$ in our study. Among all the proposed forward physics experiments, FASER$\nu2$ has the best sensitivity to the 
$U(1)_{B-L}$ and the dark photon model,
whereas the SND@LHC has the least. The sensitivity curves for all the experiments follow a similar pattern, in which the sensitivity gets
weakened with the increment of dark photon mass. 
We also compared the 90\% C.L. sensitivity reach on the parameter space of the models at the FPF experiments with the other existing bounds from the experiments such as TEXONO, GEMMA, BOREXINO, LSND, CHARM II, NA64e. 

In summary, the FPF experiments considered in this work can achieve better 
sensitivities at higher $M_{A'}$ mass region because of the higher 
energy range of the neutrino flux coming from the ATLAS IP.

Before closing we would like to address an interesting idea of testing the dark
photon model at ICECUBE. To our knowledge
the dark photon studies at ICECUBE \cite{Ardid:2017lry} 
were based on the decay of dark photons into
charged leptons or pions, followed by cascade decays into neutrinos, e.g.,
$\mu^- \to e^- \bar \nu_e \nu_\mu$ and $\pi^- \to \mu^-  \bar \nu_\mu$. 
On the other hand, using neutrino-electron or neutrino-nucleon
scattering via dark photon, one can use  the existing ICECUBE neutrino flux
measurements \cite{IceCube:2021uhz,IceCube:2020acn} to constrain the dark photon
model parameters.  The measurement of muon-neutrino flux is based on
induced muon tracks.  Electron- and tau-neutrino fluxes are determined
by electromagnetic and/or hadronic cascade caused by the electron or low-energy
tau neutrinos.  The presence of the dark photon $A'$ would modify the scattering
of $\nu e^- \to \nu e^-$, which would then affect the electron-neutrino flux
measurement.  Since the energy range of the neutrino at ICECUBE
is much higher than that of FASER$\nu(2)$, we expect ICECUBE could be 
more sensitive to higher mass region of $M_{A'} > 1$  GeV. We will explore this
possibility in the future.

\section*{Acknowledgement}
This work was supported in parts by the Ministry of Science and Technology
with the grant number MoST-110-2112-M-007-017-MY3.

\end{document}